\documentclass[
preprint,
showpacs
]{revtex4-1}
\usepackage{amsthm}
\theoremstyle{plain}
\newtheorem{theorem}{Theorem}
\newtheorem*{theorem*}{Theorem}

\newtheorem*{definition*}{Definition}

\newtheorem*{lemma*}{Lemma}

\newtheorem{remark}[theorem]{Remark}

\usepackage{braket}
\usepackage{delarray}
\usepackage{here}

\usepackage{amsmath}
\usepackage{txfonts}

\usepackage[dvipdfmx]{graphicx}
\usepackage{bm}
\usepackage{color}
\usepackage{ulem}

\usepackage{physics}

\newcommand{\ba}{\begin{array}}
\newcommand{\ea}{\end{array}}
\newcommand{\bmat}{\left(\begin{array}}
\newcommand{\emat}{\end{array}\right)}
\newcommand{\no}{\nonumber}

\newcommand{\be}{\begin{eqnarray}}
\newcommand{\ee}{\end{eqnarray}}

\begin{document}
\title{Griffiths inequalities and Gibbs-Bogoliubov inequality for general gauge glasses with Gaussian disorder on Nishimori line}
\author{Manaka Okuyama$^{1,2}$}
\author{Masayuki Ohzeki$^{1,3,4,5}$}
\affiliation{$^1$Graduate School of Information Sciences, Tohoku University, Sendai 980-8579, Japan}
\affiliation{$^2$School of Computing, Institute of Science Tokyo, Tokyo 152-8551, Japan}
\affiliation{$^3$Department of Physics, Institute of Science Tokyo, Tokyo 152-8551, Japan}
\affiliation{$^4$Research and Education Institute for Semiconductors and Informatics, Kumamoto University, Kumamoto 860-0862, Japan}
\affiliation{$^5$Sigma-i Co., Ltd., Tokyo 108-0075, Japan} 

\begin{abstract}  
We consider a class of gauge glass models with Gaussian disorder on the Nishimori line, including the Ising spin glass, the $XY$ gauge glass, the $Z_q$ gauge glass, and the gauge-invariant Potts model.
We prove that the first and second Griffiths inequalities hold for these models on arbitrary lattice structures.
As a consequence, both the pressure and the correlation functions are monotonically increasing with respect to the inverse temperature along the Nishimori line.
Furthermore, we establish an analogue of the Gibbs--Bogoliubov inequality for this class of models.
This result implies that, on the Nishimori line, the approximate quenched free energy obtained via the replica method with a replica-symmetric mean-field approximation is always greater than the true quenched free energy.
Our results provide a broad generalization of previous results established for the Ising spin glass with Gaussian disorder on the Nishimori line.
\end{abstract}
\date{\today}
\maketitle

\section{Introduction}

Correlation inequalities have played a crucial role in the study of phase transitions in statistical mechanics~\cite{FV}.
In particular, the Griffiths inequalities, also known as the Griffiths--Kelly--Sherman (GKS) inequalities~\cite{Griffiths,KS}, which hold for a class of ferromagnetic models, are powerful tools for establishing the existence of the thermodynamic limits of the pressure and correlation functions, deriving bounds on the critical temperature, and proving the existence of phase transitions.
Roughly speaking, the first Griffiths inequality states that the correlation functions are non-negative and that the pressure is a monotonically increasing function of the inverse temperature, while the second Griffiths inequality asserts that the correlation functions increase monotonically with the inverse temperature.
The Griffiths inequalities were originally proved for the Ising model and later extended to various ferromagnetic spin systems~\cite{Ginibre,KPC,Dunlop,BLU}, including the $XY$ model and the Potts model.

The derivation of the Griffiths inequalities relies crucially on the positivity of the interactions, and therefore these inequalities do not, at first glance, apply to spin glass models with both positive and negative couplings.
Over the past few decades, several attempts have been made to establish analogues of the GKS inequalities for spin glass models~\cite{CG,CL,CG2,MNC,Kitatani,OO,CUV}.
In particular, complete counterparts of the first and second Griffiths inequalities have been established for the Ising spin glass model on the Nishimori line~\cite{MNC} and have been utilized in rigorous analyses of finite-dimensional systems~\cite{CSH,OO2,OO3}.
The Nishimori line~\cite{Nishimori,Nishimori2,Nishimori3} is a special subspace in the phase diagram on which various rigorous results hold.
In particular, the quenched average of the correlation functions is non-negative on the Nishimori line.
The validity of the Griffiths inequalities for the Ising spin glass model on the Nishimori line can therefore be understood as a manifestation of its effective ferromagnetic nature.
The Nishimori line exists not only for the Ising spin glass model but also for a broader class of gauge glass models~\cite{NS,ON}, including the $XY$ gauge glass, the $Z_q$ gauge glass, and the gauge-invariant Potts model.
This naturally raises the question: do the Griffiths inequalities on the Nishimori line also hold for gauge glass models beyond the Ising spin glass model?

In the present study, we consider a general class of gauge glass models with Gaussian disorder~\cite{ON}, including the Ising spin glass, the $XY$ gauge glass, the $Z_q$ gauge glass, and the gauge-invariant Potts model.
We prove that the Griffiths inequalities hold for these models on the Nishimori line on arbitrary lattice structures.
The key ingredient of the proof is an integration-by-parts formula for the Gaussian distribution.
The proof follows essentially the same strategy as that used for the Ising spin glass model with Gaussian disorder on the Nishimori line~\cite{MNC}.

Furthermore, using the derivation of the Griffiths inequalities on the Nishimori line, we show that the pressure is convex with respect to a certain parameter along the Nishimori line.
This convexity is distinct from the conventional convexity with respect to the inverse temperature.
Exploiting this property, we prove that an analogue of the Gibbs--Bogoliubov inequality holds for this class of models on the Nishimori line.
The conventional Gibbs--Bogoliubov inequality~\cite{Kuzemsky}, which is based on convexity with respect to the inverse temperature, provides an upper bound on the true free energy in terms of a trial function.
For example, it implies that the free energy obtained within the mean-field approximation is always greater than the true free energy.
By applying this analogue of the Gibbs--Bogoliubov inequality on the Nishimori line, we show that the approximate free energy obtained via the replica method with a replica-symmetric mean-field approximation is always greater than the true free energy.
Related results have previously been obtained for the Ising spin glass model on the Nishimori line~\cite{OO4}.

The remainder of this paper is organized as follows.
In Section~II, we define a general class of gauge glass models with Gaussian disorder on the Nishimori line and state the main results.
In Section~III, we prove the main results.
In Section~IV, we conclude with a discussion of the results and related open problems.

\section{Model and results}
\subsection{Model}
Following the result by Ozeki and Nishimori~\cite{ON}, we consider a general class of gauge glass models with Gaussian disorder.
Let $\Lambda$ be a set of $N$ lattice sites, and let $\Lambda_B \subset \Lambda \otimes \Lambda$ denote a set of $N_B$ bonds, each consisting of an unordered pair of sites in $\Lambda$.
The lattice structure is arbitrary, and we impose free boundary conditions.
At each site $i \in \Lambda$, we assign a spin variable $\phi_i \in \Phi$, where $\Phi$ is equipped with a non-negative measure $d\mu(\phi_i)$.
Here, $\Phi$ denotes the space of spin states at a single lattice site.
We write $\{\phi\} = (\phi_1, \ldots, \phi_N) \in \Phi^N$ for a spin configuration.
The corresponding product measure is given by
\[
d\mu(\phi) = \prod_{i \in \Lambda} d\mu(\phi_i).
\]
Throughout this paper, we assume that $\Phi$ is a compact Abelian group; in particular, $\Phi \subseteq [0,2\pi)$.

To simplify the presentation, we focus on two-body interactions, although the results extend to one-body and many-body interactions.
We consider the random potential
\begin{align}
U = - \sum_{\langle ij \rangle \in \Lambda_B} \sum_{m}' \beta_{ij,m}
\, \Re \left[ \omega_{ij,m} \, e^{ i m (\phi_i - \phi_j)} \right].
\label{model}
\end{align}
Here, each pair $\langle ij \rangle$ is counted once, $\beta_{ij,m}$ denotes the local inverse temperature, and the primed sum $\sum_m'$ runs over integers $m$ such that $\beta_{ij,m} > 0$.
The variables $\omega_{ij,m}$ are independent complex Gaussian random variables with density
\begin{align}
P(\{\omega\})
= \prod_{\langle ij \rangle \in \Lambda_B} \prod_{m}'
\frac{1}{2\pi \sigma_{ij,m}^2}
\exp\left(
-\frac{|\omega_{ij,m} - D_{ij,m}|^2}{2\sigma_{ij,m}^2}
\right),
\end{align}
where $\sigma_{ij,m} \ge 0$ and $D_{ij,m} \ge 0$, and the density is defined with respect to the Lebesgue measure on the complex plane.
The primed product $\prod_m'$ runs over integers $m$ such that $\beta_{ij,m} > 0$.
The corresponding probability measure is given by
\begin{align}
d\nu(\{\omega\})
= P(\{\omega\}) \prod_{\langle ij \rangle \in \Lambda_B} \prod_{m}' d^2\omega_{ij,m},
\end{align}
where $d^2\omega = d(\Re \omega)\, d(\Im \omega)$.
Introducing two independent real Gaussian random variables $J_{ij,m}$ and $K_{ij,m}$, we write
\[
\omega_{ij,m} = J_{ij,m} - i K_{ij,m}.
\]
The joint distribution of $\{J,K\}$ is
\begin{align}
P(\{J,K\})
= \prod_{\langle ij \rangle \in \Lambda_B} \prod_{m}'
\frac{1}{2\pi \sigma_{ij,m}^2}
\exp\left(
- \frac{(J_{ij,m} - D_{ij,m})^2 + K_{ij,m}^2}{2\sigma_{ij,m}^2}
\right).
\end{align}
With this parametrization, the random potential becomes
\begin{align}
U
= - \sum_{\langle ij \rangle \in \Lambda_B} \sum_{m}'
\beta_{ij,m}\Bigl(
J_{ij,m} \cos\bigl(m(\phi_i - \phi_j)\bigr)
+ K_{ij,m} \sin\bigl(m(\phi_i - \phi_j)\bigr)
\Bigr).
\end{align}
The partition function is defined by
\begin{align}
Z = \int d\mu(\phi)\, e^{-U}.
\end{align}
The thermal average $\langle \cdots \rangle$ and the quenched average $\mathbb{E}[\cdots]$ are defined as
\begin{align}
\langle \cdots \rangle
&= \frac{\int d\mu(\phi)\, (\cdots)\, e^{-U}}{\int d\mu(\phi)\, e^{-U}},
\\
\mathbb{E}[\cdots]
&= \int d\nu(\{\omega\})\, (\cdots).
\end{align}
The quenched pressure is defined by
\begin{align}
P = \mathbb{E}[\log Z].
\end{align}

The model is invariant under the following gauge transformation:
\begin{align}
\phi_i \;\to\; \phi_i - \theta_i, \qquad
\omega_{ij,m} \;\to\; \omega_{ij,m} \, e^{i m (\theta_i - \theta_j)},
\end{align}
for arbitrary $\{\theta_i\}_{i \in \Lambda} \subset \Phi$.
Under this transformation, the random potential $U$ remains unchanged.
This gauge invariance plays a central role in deriving exact identities on the Nishimori line~\cite{ON}.
The Nishimori-line condition is given by
\begin{align}
\frac{D_{ij,m}}{\sigma_{ij,m}^2}
= \beta_{ij,m},
\label{NL-condition}
\end{align}
for all $\langle ij \rangle \in \Lambda_B$ and $m$.
Under this condition, various exact results follow from gauge transformations.
In particular, the following correlation identity~\cite{Nishimori2} holds for any $m$ and any pair of sites $i,j$:
\begin{align}
\mathbb{E}\!\left[ \left\langle \cos m(\phi_i - \phi_j) \right\rangle \right]
=
\mathbb{E}\!\left[
\left\langle
\cos m\bigl( (\phi_i^1 - \phi_j^1) - (\phi_i^2 - \phi_j^2) \bigr)
\right\rangle_{1,2}
\right],
\end{align}
where the variables $\phi_i^1$ and $\phi_i^2$ represent the spins at site $i$ in the first and second replicas, respectively, and $\langle \cdots \rangle_{1,2}$ denotes the thermal average over the two replicas.
This identity can be rewritten as
\begin{align}
\mathbb{E}\!\left[ \left\langle \cos m(\phi_i - \phi_j) \right\rangle \right]
=
\mathbb{E}\!\left[
\left\langle \cos m(\phi_i - \phi_j) \right\rangle^2
+
\left\langle \sin m(\phi_i - \phi_j) \right\rangle^2
\right]
\ge 0.
\label{corre-identy}
\end{align}
Thus, the quenched average of the correlation function is non-negative for any $m$, reflecting the effective ferromagnetic nature of the system along the Nishimori line.
Furthermore, several related correlation identities can be derived on the Nishimori line using gauge transformations:
\begin{align}
&\mathbb{E}\bigl[ \langle \cos m(\phi_i - \phi_j) \cos n(\phi_k - \phi_l) \rangle \bigr]
\no\\
=&
\mathbb{E}\bigl[
\langle
\cos m\bigl((\phi_i^1 - \phi_j^1) - (\phi_i^2 - \phi_j^2)\bigr)
\cos n\bigl((\phi_k^1 - \phi_l^1) - (\phi_k^2 - \phi_l^2)\bigr)
\rangle_{1,2}
\bigr],
\label{corre-identity2}
\\
&\mathbb{E}\bigl[
\langle \cos m(\phi_i - \phi_j)\rangle
\langle \cos n(\phi_k - \phi_l)\rangle
\bigr]
\no\\
=&
\mathbb{E}\bigl[
\langle
\cos m\bigl((\phi_i^1 - \phi_j^1) - (\phi_i^3 - \phi_j^3)\bigr)
\cos n\bigl((\phi_k^1 - \phi_l^1) - (\phi_k^2 - \phi_l^2)\bigr)
\rangle_{1,2,3}
\bigr],
\label{corre-identity3}
\\
&\mathbb{E}\bigl[
\langle
\cos m(\phi_i^1 - \phi_j^1)
\cos n\bigl((\phi_k^1 - \phi_l^1) - (\phi_k^2 - \phi_l^2)\bigr)
\rangle_{1,2}
\bigr]
\no\\
=&
\mathbb{E}\bigl[
\langle
\cos m\bigl((\phi_i^1 - \phi_j^1) - (\phi_i^3 - \phi_j^3)\bigr)
\cos n\bigl((\phi_k^1 - \phi_l^1) - (\phi_k^2 - \phi_l^2)\bigr)
\rangle_{1,2,3}
\bigr],
\label{corre-identity4}
\\
&\mathbb{E}\bigl[
\langle
\cos m(\phi_i^1 - \phi_j^1)
\cos n\bigl((\phi_k^2 - \phi_l^2) - (\phi_k^3 - \phi_l^3)\bigr)
\rangle_{1,2,3}
\bigr]
\no\\
=&
\mathbb{E}\bigl[
\langle
\cos m\bigl((\phi_i^1 - \phi_j^1) - (\phi_i^4 - \phi_j^4)\bigr)
\cos n\bigl((\phi_k^2 - \phi_l^2) - (\phi_k^3 - \phi_l^3)\bigr)
\rangle_{1,2,3,4}
\bigr].
\label{corre-identity5}
\end{align}
Here, $\langle \cdots \rangle_{1,2,3}$ and $\langle \cdots \rangle_{1,2,3,4}$ denote the thermal averages over three and four replicas, respectively.
The variables $\phi_i^3$ and $\phi_i^4$ represent the spins at site $i$ in the third and fourth replicas.
These identities play a crucial role in the following analysis.

\subsection{Examples}
By appropriate choices of the spin variables $\phi_i$, the model~\eqref{model} reduces to several well-known models~\cite{ON}.
\paragraph{Ising spin glass.}
By choosing $\Phi = \{0,\pi\}$, $\sigma_i \equiv e^{i\phi_i}$, and retaining only the $m=1$ term, the model reduces to the Ising spin glass:
\begin{align}
U
= - \sum_{\langle ij \rangle \in \Lambda_B}
\beta_{ij} J_{ij} \sigma_i \sigma_j ,
\end{align}
where the variables $K_{ij}$ do not contribute.

\paragraph{$XY$ gauge glass.}
By choosing $\Phi = [0,2\pi)$ and retaining only the $m=1$ term, the model reduces to the $XY$ gauge glass:
\begin{align}
U
= - \sum_{\langle ij \rangle \in \Lambda_B}
\beta_{ij}
\bigl(
J_{ij} \cos(\phi_i - \phi_j)
+ K_{ij} \sin(\phi_i - \phi_j)
\bigr).
\end{align}

\paragraph{$Z_q$ gauge glass.}
Similarly, by choosing
\[
\Phi =
\left\{
0, \frac{2\pi}{q}, \frac{4\pi}{q}, \ldots, \frac{2(q-1)\pi}{q}
\right\},
\]
with a positive integer $q$, and retaining only the $m=1$ term, the model reduces to the $Z_q$ gauge glass:
\begin{align}
U
= - \sum_{\langle ij \rangle \in \Lambda_B}
\beta_{ij}
\bigl(
J_{ij} \cos(\phi_i - \phi_j)
+ K_{ij} \sin(\phi_i - \phi_j)
\bigr).
\end{align}

\paragraph{Gauge-invariant Potts model.}
Finally, by choosing
\[
\Phi =
\left\{
0, \frac{2\pi}{q}, \frac{4\pi}{q}, \ldots, \frac{2(q-1)\pi}{q}
\right\},
\qquad
\lambda_i \equiv e^{i\phi_i},
\]
the model reduces to the gauge-invariant Potts model:
\begin{align}
U
= - \sum_{\langle ij \rangle \in \Lambda_B}
\sum_{m=0}^{q-1}
\beta_{ij,m}
\, \Re\!\left[
\omega_{ij,m} \lambda_i^m \lambda_j^{q-m}
\right].
\end{align}

\subsection{Main Results}
We introduce a parameter on the Nishimori line~\cite{MNC}:
\begin{align}
x_{ij,m}
= \beta_{ij,m} D_{ij,m}
= \sigma_{ij,m}^2 \beta_{ij,m}^2 .
\end{align}
The limit $x_{ij,m} \to 0$ corresponds to the high-temperature regime with zero mean, whereas $x_{ij,m} \to \infty$ corresponds to the low-temperature regime with a strong ferromagnetic bias.
Thus, by varying $x_{ij,m}$, one can continuously interpolate between high and low temperatures along the Nishimori line.

We now state the Griffiths inequalities for general gauge glass models with Gaussian disorder on the Nishimori line.
\begin{theorem}[First Griffiths inequality]
\label{first-Griffiths}
For any choice of parameters $\{x_{ij,m}\}$ on the Nishimori line, the quenched pressure $\mathbb{E}[\log Z]$ is monotonically increasing with respect to each $x_{ij,m}$:
\begin{align}
\frac{\partial}{\partial x_{ij,m}} \mathbb{E}[\log Z]
= \frac{1}{2} \, \mathbb{E}\!\left[ 1 + \left\langle \cos m(\phi_i - \phi_j) \right\rangle \right]
\ge 0 .
\end{align}
\end{theorem}

\begin{theorem}[Second Griffiths inequality]
\label{second-Griffiths}
For any two pairs $(ij,m)$ and $(kl,n)$, the quenched pressure satisfies
\begin{align}
&2 \frac{\partial^2}{\partial x_{ij,m} \, \partial x_{kl,n}}
\mathbb{E}[\log Z]
\nonumber\\
&= \frac{\partial}{\partial x_{kl,n}}
\mathbb{E}\!\left[ \left\langle \cos m(\phi_i - \phi_j) \right\rangle \right]
\nonumber\\
&= \mathbb{E}\!\left[
\Bigl(
\langle \cos m(\phi_i - \phi_j)
\cos n(\phi_k - \phi_l) \rangle
-
\langle \cos m(\phi_i - \phi_j) \rangle
\langle \cos n(\phi_k - \phi_l) \rangle
\Bigr)^2
\right]
\nonumber\\
&\quad
+ \mathbb{E}\!\left[
\Bigl(
\langle \cos m(\phi_i - \phi_j)
\sin n(\phi_k - \phi_l) \rangle
-
\langle \cos m(\phi_i - \phi_j) \rangle
\langle \sin n(\phi_k - \phi_l) \rangle
\Bigr)^2
\right]
\nonumber\\
&\quad
+ \mathbb{E}\!\left[
\Bigl(
\langle \sin m(\phi_i - \phi_j)
\cos n(\phi_k - \phi_l) \rangle
-
\langle \sin m(\phi_i - \phi_j) \rangle
\langle \cos n(\phi_k - \phi_l) \rangle
\Bigr)^2
\right]
\nonumber\\
&\quad
+ \mathbb{E}\!\left[
\Bigl(
\langle \sin m(\phi_i - \phi_j)
\sin n(\phi_k - \phi_l) \rangle
-
\langle \sin m(\phi_i - \phi_j) \rangle
\langle \sin n(\phi_k - \phi_l) \rangle
\Bigr)^2
\right]
\ge 0 .
\end{align}
\end{theorem}

Furthermore, a simple rearrangement of the expression for the second derivatives in Theorem~\ref{second-Griffiths} shows that the quenched pressure is convex on the Nishimori line.
\begin{theorem}[Convexity on the Nishimori line]
\label{convexity-nishimori}
The quenched pressure $\mathbb{E}[\log Z]$ is convex with respect to the parameters $\{x_{ij,m}\}$.
\end{theorem}

\begin{remark}
This convexity is distinct from the conventional convexity of the pressure with respect to the inverse temperature.
\end{remark}

\begin{remark}
Theorems~\ref{first-Griffiths}, \ref{second-Griffiths}, and \ref{convexity-nishimori}
were previously established for the Ising spin glass on the Nishimori line~\cite{MNC,OO4}.
In that case, they reduce to
\begin{align}
\frac{\partial}{\partial x_{ij}}
\mathbb{E}[\log Z]
&= \frac{1}{2} \, \mathbb{E}\!\left[ 1 + \left\langle \sigma_i \sigma_j \right\rangle \right]
\ge 0 ,
\end{align}
and
\begin{align}
2 \frac{\partial^2}{\partial x_{ij} \, \partial x_{kl}}
\mathbb{E}[\log Z]
&= \frac{\partial}{\partial x_{kl}}
\mathbb{E}\!\left[ \left\langle \sigma_i \sigma_j \right\rangle \right]
\nonumber\\
&= \mathbb{E}\!\left[
\Bigl(
\langle \sigma_i \sigma_j \sigma_k \sigma_l \rangle
-
\langle \sigma_i \sigma_j \rangle
\langle \sigma_k \sigma_l \rangle
\Bigr)^2
\right]
\ge 0 .
\end{align}
\end{remark}

\begin{remark}
Theorems~\ref{first-Griffiths}, \ref{second-Griffiths}, and~\ref{convexity-nishimori}
remain valid even in the presence of one-body and many-body interactions.
For example, the following interaction terms can be included in the random potential~\eqref{model}:
\begin{align}
\Re\!\left[ \beta_{i,m} \, \omega_{i,m} \, e^{ i m \phi_i } \right],
\qquad
\Re\!\left[ \beta_{ijk,m} \, \omega_{ijk,m} \,
e^{ i m (\phi_i - \phi_j - \phi_k) } \right].
\end{align}
\end{remark}

Finally, by exploiting the convexity of the pressure on the Nishimori line,
we derive an analogue of the Gibbs--Bogoliubov (GB) inequality~\cite{OO4}.
This inequality provides a lower bound on the pressure in terms of a suitable trial potential.
To simplify the presentation, we consider the case in which the pressure associated with the two-body random potential~\eqref{model} is approximated by a one-body trial potential.
We introduce the one-body random potential
\begin{align}
U_0
&= - \sum_{i \in \Lambda} \sum_{m}'
\beta_{i,m}
\Bigl(
J_{i,m} \cos(m\phi_i)
+ K_{i,m} \sin(m\phi_i)
\Bigr),
\end{align}
where $\beta_{i,m}$ denotes the local inverse temperature, and $J_{i,m}$ and $K_{i,m}$ are quenched random variables independently drawn from Gaussian distributions with variance $\sigma_{i,m}^2$ and means $D_{i,m}$ and $0$, respectively.
We impose the Nishimori-line condition
\begin{align}
x_{i,m} = \beta_{i,m} D_{i,m} = \sigma_{i,m}^2 \beta_{i,m}^2 .
\end{align}
The corresponding partition function is defined by
\begin{align}
Z_0 = \int d\mu(\phi)\, e^{-U_0}.
\end{align}
The thermal average with respect to $U_0$ is defined as
\begin{align}
\langle \cdots \rangle_0
= \frac{\int d\mu(\phi)\, (\cdots)\, e^{-U_0}}
{\int d\mu(\phi)\, e^{-U_0}} .
\end{align}
We are now in a position to state the following result.

\begin{theorem}[Gibbs--Bogoliubov inequality on the Nishimori line]
\label{GB-NL}
For any sets of parameters $\{x_{ij,m}\}$ and $\{x_{i,m}\}$,
the quenched pressure $\mathbb{E}[\log Z]$ satisfies
\begin{align}
\mathbb{E}[\log Z]
&\ge \mathbb{E}[\log Z_0]
+ \frac{1}{2} \sum_{\langle ij \rangle \in \Lambda_B} \sum_{m}' x_{ij,m}\, 
\mathbb{E}\!\left[ 1 + \left\langle \cos m(\phi_i - \phi_j) \right\rangle_0 \right]
\nonumber\\
&\quad
- \frac{1}{2} \sum_{i \in \Lambda} \sum_{m}' x_{i,m}\,
\mathbb{E}\!\left[ 1 + \left\langle \cos(m \phi_i) \right\rangle_0 \right].
\end{align}
\end{theorem}

\begin{remark}
As an illustration, consider the Ising spin glass model with coordination number $z$ and periodic boundary conditions.
By setting $x_{ij} = \beta^2$ and $x_i = \beta^2 z M$, and maximizing the right-hand side with respect to $M$,
one finds that the resulting bound coincides with that obtained via the replica method with a replica-symmetric mean-field approximation on the Nishimori line~\cite{OO4}.
This implies that, on the Nishimori line, the free energy obtained via the replica method followed by a mean-field approximation is always greater than the true free energy.
This corresponds to the well-known fact that, in conventional ferromagnetic systems, the mean-field approximation provides an upper bound on the true free energy.
\end{remark}

\section{Proof of main results}
In this section, for notational simplicity, we rescale the random variables
$J_{ij,m}$ and $K_{ij,m}$ so that they follow standard Gaussian distributions with mean zero and unit variance.
More precisely, after an appropriate change of variables, the random potential can be written as
\begin{align}
U
= - \sum_{\langle ij \rangle \in \Lambda_B} \sum_{m}'
\Bigl(
\sqrt{x_{ij,m}}\, J_{ij,m} \cos m(\phi_i - \phi_j)
+ \sqrt{x_{ij,m}}\, K_{ij,m} \sin m(\phi_i - \phi_j)
+ x_{ij,m} \cos m(\phi_i - \phi_j)
\Bigr),
\end{align}
and the joint probability distribution of $\{J,K\}$ is given by
\begin{align}
P(\{J,K\})
= \prod_{\langle ij \rangle \in \Lambda_B} \prod_{m}'
\frac{1}{2\pi}
\exp\left(
- \frac{J_{ij,m}^2 + K_{ij,m}^2}{2}
\right).
\end{align}

\subsection{Proof of Theorem~\ref{first-Griffiths}}
Using Gaussian integration by parts, we obtain
\begin{align}
\frac{\partial}{\partial x_{ij,m}} \mathbb{E}[\log Z]
&=
\frac{1}{2\sqrt{x_{ij,m}}}
\mathbb{E}\!\left[
J_{ij,m} \langle \cos m(\phi_i - \phi_j) \rangle
+ K_{ij,m} \langle \sin m(\phi_i - \phi_j) \rangle
\right]
\nonumber\\
&\quad
+ \mathbb{E}\!\left[
\langle \cos m(\phi_i - \phi_j) \rangle
\right]
\nonumber\\
&=
\frac{1}{2}
\mathbb{E}\!\left[
\langle \cos^2 m(\phi_i - \phi_j) \rangle
- \langle \cos m(\phi_i - \phi_j) \rangle^2
\right]
\nonumber\\
&\quad
+ \frac{1}{2}
\mathbb{E}\!\left[
\langle \sin^2 m(\phi_i - \phi_j) \rangle
- \langle \sin m(\phi_i - \phi_j) \rangle^2
\right]
\nonumber\\
&\quad
+ \mathbb{E}\!\left[
\langle \cos m(\phi_i - \phi_j) \rangle
\right]
\nonumber\\
&=
\frac{1}{2}
\mathbb{E}\!\left[
1
- \langle \cos m(\phi_i - \phi_j) \rangle^2
- \langle \sin m(\phi_i - \phi_j) \rangle^2
\right]
\nonumber\\
&\quad
+ \mathbb{E}\!\left[
\langle \cos m(\phi_i - \phi_j) \rangle
\right].
\end{align}
Using the identity~\eqref{corre-identy}, we further obtain
\begin{align}
\frac{\partial}{\partial x_{ij,m}} \mathbb{E}[\log Z]
&=
\frac{1}{2}
\mathbb{E}\!\left[
1 - \langle \cos m(\phi_i - \phi_j) \rangle
\right]
+ \mathbb{E}\!\left[
\langle \cos m(\phi_i - \phi_j) \rangle
\right]
\nonumber\\
&=
\frac{1}{2}
\mathbb{E}\!\left[
1 + \langle \cos m(\phi_i - \phi_j) \rangle
\right]
\nonumber\\
&\ge 0.
\end{align}
This completes the proof of Theorem~\ref{first-Griffiths}.

\subsection{Proof of Theorem~\ref{second-Griffiths}}
We begin by differentiating the correlation function with respect to $x_{kl,n}$.
A direct computation yields
\begin{align}
&\frac{\partial}{\partial x_{kl,n}} \mathbb{E}\bigl[ \langle \cos m(\phi_i - \phi_j)\rangle \bigr]
\nonumber\\
&=
\frac{1}{2\sqrt{x_{kl,n}}}
\mathbb{E}\!\left[
J_{kl,n}\langle \cos m(\phi_i - \phi_j)\cos n(\phi_k - \phi_l)\rangle
- J_{kl,n}\langle \cos m(\phi_i - \phi_j)\rangle
\langle \cos n(\phi_k - \phi_l)\rangle
\right]
\nonumber\\
&\quad
+ \frac{1}{2\sqrt{x_{kl,n}}}
\mathbb{E}\!\left[
K_{kl,n}\langle \cos m(\phi_i - \phi_j)\sin n(\phi_k - \phi_l)\rangle
- K_{kl,n}\langle \cos m(\phi_i - \phi_j)\rangle
\langle \sin n(\phi_k - \phi_l)\rangle
\right]
\nonumber\\
&\quad
+ \mathbb{E}\!\left[
\langle \cos m(\phi_i - \phi_j)\cos n(\phi_k - \phi_l)\rangle
- \langle \cos m(\phi_i - \phi_j)\rangle
\langle \cos n(\phi_k - \phi_l)\rangle
\right].
\label{corr-deri}
\end{align}
We next apply Gaussian integration by parts.
The first term in Eq.~\eqref{corr-deri} becomes
\begin{align}
\frac{1}{2\sqrt{x_{kl,n}}}\mathbb{E}[J_{kl,n}\cdots]
=
\frac{1}{2}\mathbb{E}\!\left[
-2 \langle \cos m(\phi_i - \phi_j)\cos n(\phi_k - \phi_l)\rangle
\langle \cos n(\phi_k - \phi_l)\rangle
\right.
\nonumber\\
\left.
+ 2 \langle \cos m(\phi_i - \phi_j)\rangle
\langle \cos n(\phi_k - \phi_l)\rangle^2
\right],
\label{first-term}
\end{align}
and similarly, the second term reduces to
\begin{align}
\frac{1}{2\sqrt{x_{kl,n}}}\mathbb{E}[K_{kl,n}\cdots]
=
\frac{1}{2}\mathbb{E}\!\left[
-2 \langle \cos m(\phi_i - \phi_j)\sin n(\phi_k - \phi_l)\rangle
\langle \sin n(\phi_k - \phi_l)\rangle
\right.
\nonumber\\
\left.
+ 2 \langle \cos m(\phi_i - \phi_j)\rangle
\langle \sin n(\phi_k - \phi_l)\rangle^2
\right].
\label{second-term}
\end{align}
Combining Eqs.~\eqref{first-term} and~\eqref{second-term}, the first two contributions in Eq.~\eqref{corr-deri} can be written as
\begin{align}
&\mathbb{E}\!\left[
- \langle \cos m(\phi_i^1-\phi_j^1)
\cos n\bigl((\phi_k^1-\phi_l^1)-(\phi_k^2-\phi_l^2)\bigr)\rangle_{1,2}
\right]
\nonumber\\
&\quad
+ \mathbb{E}\!\left[
\langle \cos m(\phi_i^1-\phi_j^1)
\cos n\bigl((\phi_k^2-\phi_l^2)-(\phi_k^3-\phi_l^3)\bigr)\rangle_{1,2,3}
\right].
\end{align}
Using the correlation identities~\eqref{corre-identity4} and~\eqref{corre-identity5}, this expression becomes
\begin{align}
&\mathbb{E}\!\left[
- \langle \cos m\bigl((\phi_i^1-\phi_j^1)-(\phi_i^3-\phi_j^3)\bigr)
\cos n\bigl((\phi_k^1-\phi_l^1)-(\phi_k^2-\phi_l^2)\bigr)\rangle_{1,2,3}
\right]
\nonumber\\
&\quad
+ \mathbb{E}\!\left[
\langle \cos m\bigl((\phi_i^1-\phi_j^1)-(\phi_i^4-\phi_j^4)\bigr)
\cos n\bigl((\phi_k^2-\phi_l^2)-(\phi_k^3-\phi_l^3)\bigr)\rangle_{1,2,3,4}
\right].
\label{first-and-second-term}
\end{align}
On the other hand, using Eqs. ~\eqref{corre-identity2} and~\eqref{corre-identity3}, the third term in Eq.~\eqref{corr-deri} becomes
\begin{align}
&\mathbb{E}\!\left[
\langle \cos m(\phi_i-\phi_j)\cos n(\phi_k-\phi_l)\rangle
- \langle \cos m(\phi_i-\phi_j)\rangle
\langle \cos n(\phi_k-\phi_l)\rangle
\right]
\nonumber\\
&=
\mathbb{E}\!\left[
\langle \cos m\bigl((\phi_i^1-\phi_j^1)-(\phi_i^2-\phi_j^2)\bigr)
\cos n\bigl((\phi_k^1-\phi_l^1)-(\phi_k^2-\phi_l^2)\bigr)\rangle_{1,2}
\right]
\nonumber\\
&\quad
-
\mathbb{E}\!\left[
\langle \cos m\bigl((\phi_i^1-\phi_j^1)-(\phi_i^3-\phi_j^3)\bigr)
\cos n\bigl((\phi_k^1-\phi_l^1)-(\phi_k^2-\phi_l^2)\bigr)\rangle_{1,2,3}
\right].
\label{third-term}
\end{align}
Substituting Eqs.~\eqref{first-and-second-term} and~\eqref{third-term} into Eq.~\eqref{corr-deri} and rearranging terms, we obtain
\begin{align}
&\frac{\partial}{\partial x_{kl,n}} \mathbb{E}\bigl[ \langle \cos m(\phi_i-\phi_j)\rangle \bigr]
\nonumber\\
&=
\mathbb{E}\!\left[
\bigl(
\langle \cos m(\phi_i-\phi_j)\cos n(\phi_k-\phi_l)\rangle
-
\langle \cos m(\phi_i-\phi_j)\rangle
\langle \cos n(\phi_k-\phi_l)\rangle
\bigr)^2
\right]
\nonumber\\
&\quad
+ \mathbb{E}\!\left[
\bigl(
\langle \cos m(\phi_i-\phi_j)\sin n(\phi_k-\phi_l)\rangle
-
\langle \cos m(\phi_i-\phi_j)\rangle
\langle \sin n(\phi_k-\phi_l)\rangle
\bigr)^2
\right]
\nonumber\\
&\quad
+ \mathbb{E}\!\left[
\bigl(
\langle \sin m(\phi_i-\phi_j)\cos n(\phi_k-\phi_l)\rangle
-
\langle \sin m(\phi_i-\phi_j)\rangle
\langle \cos n(\phi_k-\phi_l)\rangle
\bigr)^2
\right]
\nonumber\\
&\quad
+ \mathbb{E}\!\left[
\bigl(
\langle \sin m(\phi_i-\phi_j)\sin n(\phi_k-\phi_l)\rangle
-
\langle \sin m(\phi_i-\phi_j)\rangle
\langle \sin n(\phi_k-\phi_l)\rangle
\bigr)^2
\right].
\end{align}
This completes the proof of Theorem~\ref{second-Griffiths}.

\subsection{Proof of Theorem~\ref{convexity-nishimori}}
The following identity plays a key role:
\begin{align}
&\Bigl(
\langle \cos m(\phi_i-\phi_j)\cos n(\phi_k-\phi_l) \rangle
- \langle \cos m(\phi_i-\phi_j)\rangle
\langle \cos n(\phi_k-\phi_l)\rangle
\Bigr)^2
\nonumber\\
&=
\Bigl\langle
\bigl(\cos m(\phi_i^1-\phi_j^1) - \langle \cos m(\phi_i-\phi_j) \rangle \bigr)
\bigl(\cos n(\phi_k^1-\phi_l^1) - \langle \cos n(\phi_k-\phi_l) \rangle \bigr)
\nonumber\\
&\qquad\quad
\bigl(\cos m(\phi_i^2-\phi_j^2) - \langle \cos m(\phi_i-\phi_j) \rangle \bigr)
\bigl(\cos n(\phi_k^2-\phi_l^2) - \langle \cos n(\phi_k-\phi_l) \rangle \bigr)
\Bigr\rangle_{1,2}.
\end{align}
We introduce the auxiliary variables
\begin{align}
c_{ij,m}
&=
\bigl(\cos m(\phi_i^1-\phi_j^1) - \langle \cos m(\phi_i-\phi_j) \rangle \bigr)
\bigl(\cos m(\phi_i^2-\phi_j^2) - \langle \cos m(\phi_i-\phi_j) \rangle \bigr),
\\
s_{ij,m}
&=
\bigl(\sin m(\phi_i^1-\phi_j^1) - \langle \sin m(\phi_i-\phi_j) \rangle \bigr)
\bigl(\sin m(\phi_i^2-\phi_j^2) - \langle \sin m(\phi_i-\phi_j) \rangle \bigr).
\end{align}
Using these variables, we obtain
\begin{align}
\frac{\partial^2}{\partial x_{ij,m}\partial x_{kl,n}}\mathbb{E}[\log Z]
&=
\frac{1}{2}
\mathbb{E}\!\left[
\langle c_{ij,m} c_{kl,n}
+ c_{ij,m} s_{kl,n}
+ s_{ij,m} c_{kl,n}
+ s_{ij,m} s_{kl,n}
\rangle
\right]
\nonumber\\
&=
\frac{1}{2}
\mathbb{E}\!\left[
\langle (c_{ij,m}+s_{ij,m})(c_{kl,n}+s_{kl,n}) \rangle
\right].
\end{align}
This shows that the Hessian matrix of $\mathbb{E}[\log Z]$ with respect to the parameters $\{x_{ij,m}\}$ is positive semidefinite.
Hence, the quenched pressure is convex on the Nishimori line.

\subsection{Proof of Theorem~\ref{GB-NL}}
We introduce the interpolating random potential
\begin{align}
U(t)
&=
- \sum_{\langle ij\rangle} \sum_m'
\Bigl(
\sqrt{t x_{ij,m}}\, J_{ij,m} \cos m(\phi_i-\phi_j)
+ \sqrt{t x_{ij,m}}\, K_{ij,m} \sin m(\phi_i-\phi_j)
+ t x_{ij,m}\cos m(\phi_i-\phi_j)
\Bigr)
\nonumber\\
&\quad
- \sum_{i \in \Lambda} \sum_m'
\Bigl(
\sqrt{(1-t) x_{i,m}}\, J_{i,m} \cos(m\phi_i)
+ \sqrt{(1-t) x_{i,m}}\, K_{i,m} \sin(m\phi_i)
+ (1-t)x_{i,m}\cos(m\phi_i)
\Bigr),
\end{align}
where $0 \le t \le 1$, and all random variables are independent standard Gaussian random variables.
The corresponding quenched pressure is
\begin{align}
\mathbb{E}[\log Z(t)], \qquad
Z(t)= \int d\mu(\phi)\, e^{-U(t)}.
\end{align}
Note that
\begin{align}
\mathbb{E}[\log Z(1)] = \mathbb{E}[\log Z], \qquad
\mathbb{E}[\log Z(0)] = \mathbb{E}[\log Z_0].
\end{align}
By Theorem~\ref{convexity-nishimori}, we have
\begin{align}
\frac{d^2}{dt^2}\mathbb{E}[\log Z(t)] \ge 0.
\end{align}
Therefore,
\begin{align}
\mathbb{E}[\log Z(1)]
\ge
\mathbb{E}[\log Z(0)]
+
\left.\frac{d}{dt}\mathbb{E}[\log Z(t)]\right|_{t=0}.
\end{align}
Applying Gaussian integration by parts, we obtain
\begin{align}
\mathbb{E}[\log Z]
&\ge \mathbb{E}[\log Z_0]
+ \frac{1}{2}\sum_{\langle ij \rangle \in \Lambda_B}\sum_m' x_{ij,m}
\mathbb{E}\!\left[
1 + \langle \cos m(\phi_i-\phi_j)\rangle_0
\right]
\nonumber\\
&\quad
- \frac{1}{2}\sum_{i \in \Lambda}\sum_m' x_{i,m}
\mathbb{E}\!\left[
1 + \langle \cos(m\phi_i)\rangle_0
\right].
\end{align}
This completes the proof of Theorem~\ref{GB-NL}.

\section{Conclusion}
Previous studies~\cite{MNC,Kitatani} have established that the Griffiths inequalities, which hold for various ferromagnetic spin systems, also hold for the Ising spin glass on the Nishimori line.
In this work, we have shown that the Griffiths inequalities can be extended to a general class of gauge glass models with Gaussian disorder on the Nishimori line.
Furthermore, we have established the convexity of the quenched pressure with respect to the parameters along the Nishimori line, which differs from the conventional convexity with respect to the inverse temperature.
Exploiting this convexity, we have derived an analogue of the Gibbs--Bogoliubov inequality on the Nishimori line.
This result implies that the quenched free energy obtained via the replica method with a replica-symmetric mean-field approximation is always greater than the true quenched free energy.

In this work, we have restricted our analysis to Gaussian disorder, for which the argument is greatly simplified by integration by parts together with correlation identities. However, the Nishimori line in gauge glass models exists even for non-Gaussian disorder~\cite{ON}, and it remains an interesting open problem whether the Griffiths inequalities continue to hold in that setting. 
For the Ising spin glass, the corresponding extension to non-Gaussian disorder was achieved by Kitatani~\cite{Kitatani}, whose argument crucially relies on special features of Ising spins, most notably the particularly simple structure of the high-temperature expansion.
By contrast, in general gauge glass models the high-temperature expansion lacks such a simple form, making Kitatani's approach difficult to extend to non-Gaussian disorder.

\section*{Acknowledgment}
This work was supported by JST BOOST, Japan Grant Number JPMJBY24B6.
This work was also supported by JSPS KAKENHI Grant Nos. 24K16973 and 23H01432.
In addition, this work was supported by programs for bridging the gap between R\&D and IDeal society (Society 5.0) and Generating Economic and social value (BRIDGE)
and Cross-ministerial Strategic Innovation Promotion Program (SIP) from the Cabinet Office (No. 23836436).

\end{document}